\documentclass[12pt]{article}

\usepackage{amsfonts}
\usepackage{amssymb}
\usepackage{amsthm}
\usepackage[fleqn]{amsmath}
\usepackage[mathcal]{euscript}
\usepackage{mathrsfs}

\newtheorem{lemma}{Lemma}
\newtheorem{proposition}{Proposition}
\newtheorem{theorem}{Theorem}

\renewcommand{\iint}{{\int\!\!\!\!\int}}

\newcommand{\Ref}[1]{(\ref{#1})}

\newcommand{\Prob}{\mathrm{Prob}\,}

\newcommand{\kB}{k_{\mathrm{B}}} 

\newcommand{\eps}{\epsilon}
\newcommand{\veps}{\varepsilon}
\newcommand{\vepskin}{\varepsilon_{\mathrm{kin}}}

\newcommand{\vect}[1] {\boldsymbol{{ #1}} }

\newcommand{\eV}{{\vect{e}}}           
\newcommand{\lV}{{\vect{l}}}           
\newcommand{\pV}{{\vect{p}}}           
\newcommand{\qV}{{\vect{q}}}           
\newcommand{\uV}{{\vect{u}}}           

\newcommand{\omegaV}{{\vect{\omega}}}  
\newcommand{\lambdaV}{{\vect{\lambda}}}

\newcommand{\pVi}{{\vect{p}_i}}        
\newcommand{\qVi}{{\vect{q}_i}}        

\newcommand{\cVN}{{\vect{C}}}          
\newcommand{\jVN}{{\vect{J}}}          
\newcommand{\lVN}{{\vect{L}}}          
\newcommand{\pVN}{{\vect{P}}}          

\newcommand{\zeroV}{\mathbf{0}}		

\DeclareMathAlphabet{\mathpzc}{OT1}{pzc}{m}{it}

\newcommand\pzcC{{\mathpzc{C}}}

\newcommand\pzcE{{\mathpzc{E}}}

\newcommand\pzcI{{\mathpzc{I}}}

\newcommand\pzcJ{{\mathpzc{J}}}

\newcommand\pzcN{{\mathpzc{N}}}

\newcommand\pzcP{{\mathpzc{P}}}

\newcommand\pzcQ{{\mathpzc{Q}}}

\newcommand\pzcS{{\mathpzc{S}}}

\newcommand\pzcPV{\vec{\pzcP}}

\newcommand\pzcQV{\vec{\pzcQ}}
\newcommand\pzcJV{\vec{\pzcJ}}

\newcommand{\abs}[1]{\left| #1 \right|}

\newcommand{\ol}[1]{\overline #1 }
\newcommand{\ul}[1]{\underline #1 }

\newcommand{\dKR}{d_{\mathrm{KR}}}

\newcommand{\dd}{\mathrm{d}}

\newcommand{\pdp}{\partial_p}

\newcommand{\pdt}{\partial_t}

\newcommand{\Eset}{\mathbb{E}}

\newcommand{\Mset}{\mathbb{M}}

\newcommand{\Rset}{\mathbb{R}}

\newcommand{\Asp}{\mathfrak{A}}

\newcommand{\Csp}{\mathfrak{C}}

\newcommand{\Lsp}{\mathfrak{L}}

\newcommand{\Tsp}{\mathfrak{T}}

\newcommand{\cP}{{\cal P}}
\newcommand{\cQ}{{\cal Q}}
\newcommand{\cX}{{\cal X}}

\newcommand{\ME}{\mathscr{ME}}

\begin{document}

\title{Statistical equilibrium dynamics}

\vspace{-0.3cm}
\author{
\normalsize \sc{Michael Kiessling}\\[-0.1cm]
\normalsize Department of Mathematics, Rutgers University\\[-0.1cm]
\normalsize Piscataway NJ 08854, USA}
\vspace{-0.3cm}
\date{Version of Nov.29, 2007.\\ Minor slips of pen corrected on May 03, 2008}
\maketitle
\vspace{-0.6cm}


\begin{abstract}
  We study the mean-field thermodynamic limit for a class
of isolated Newtonian N-body systems whose Hamiltonian admits several
additional integrals of motion. Examples are systems which are
isomorphic to plasma models consisting of one specie of charged particles
moving in a neutralizing uniform background charge. We find that in the
limit of infinitely many particles the stationary ensemble measures with
prescribed values of the integrals of motion are supported on the set of
maximum entropy solutions of a (time-independent) nonlinear
fixed point equation of mean-field type. Each maximum entropy solution of
this fixed point equation can in turn be either a static or a stationary
solution for the entropy-conserving Vlasov evolution, or even belong to a
one-dimensional orbit of maximum entropy solutions which evolve into one
another by the Vlasov dynamics. In short, the macrostates of individual
members of an equilibrium ensemble are not necessarily themselves in a
state of global statistical equilibrium in the strict sense. Yet they
are always locally in thermodynamic equilibrium, and always global
maximizers of the pertinent maximum entropy principle.
\end{abstract}

\vfill

\hrule
\smallskip
\noindent
{\small
\copyright 2007/2008 The author. 
This preprint may be reproduced by any means for noncommercial purposes.\\
(It is based on an invited talk delivered at the workshop ``Dynamics and 
Thermodynamics of Systems with Long Range Interactions: Theory and Experiments,'' Assisi, Italy, 2007.
An AIP layout of this preprint appeared in AIP Conf. Proc. \textbf{970}, pp. 91--108 (2008),
A. Campa, A. Giansanti, G. Morigi, F. Sylos Labini  (Eds.). The copyright for the AIP version
has been transferred to the AIP.)}

%
%
%
%
\section{Introduction}

  A short while ago I had an epiphany regarding equilibrium statistical mechanics.
  It happened while collaborating with Carlo Lancellotti on some curious Hamiltonian systems 
with unusual many body dynamics discovered by the Lynden-Bells \cite{LBLBa,LBLBb}.
  It could (and should!) have happened much earlier, though it wouldn't have been 
an epiphany, then, if I had understood right from the beginning what equilibrium statistical 
mechanics is all about.
  In any case, until that moment of revelation I was under the erroneous --- yet I believe 
it's fair to say: common --- impression that equilibrium statistical mechanics is entirely 
about the statistical microscopic foundations of the macroscopic concept of thermal equilibrium
for Hamiltonian $N$-body systems when $N\gg 1$. 
  Of course, equilibrium statistical mechanics is {\it also} about this, and even was invented
specifically for this purpose, but that's not the whole story.
  The complete one is much richer and much more interesting.

  In the following, I will first recall the ``folklore'' 
about Boltzmann's \cite{Boltzmann} ergodic ensemble measure, more familiar under
the name Gibbs' \cite{Gibbs} microcanonical ensemble measure
\begin{equation}
\dd\mu= \frac{1}{\Phi^{(N)}(E)} \delta(H-E)\dd^{3N}\!p\,\dd^{3N}\!q,
\label{eq:BOLTZMANNsERGODEstandard}
\end{equation}
where
\begin{equation}
\Phi^{(N)}(E) = \int \delta(H-E)\dd^{3N}\!p\,\dd^{3N}\!q
\label{eq:ZUSTANDSSUMME}
\end{equation}
is the normalizing factor. 
  I will point out misconceptions in this folklore which become particularly
troublesome  when considering the restrictions of \Ref{eq:BOLTZMANNsERGODEstandard} 
to its stationary ergodic sub-ensembles associated with additional conserved quantities
beside the Hamiltonian $H$.
  The true significance of \Ref{eq:BOLTZMANNsERGODEstandard} and its stationary sub-ensembles 
will become crystal clear once we pay due attention not just to the measures but also to the
flow with respect to which they are stationary.

 Throughout this presentation, with the exception of a few more general results it is always 
understood that the Hamiltonian $H$ of the classical $N$ body system under consideration is 
stable [i.e., collapse of parts of the system to a point is excluded] and confining [i.e., 
escape of particles from the system is excluded].
  Prime examples of such systems with long range interactions are one-component Coulomb 
systems in a neutralizing background, which in particular includes the Lynden-Bells' system 
(a 4D Coulomb system restricted to 3D motions) but also 2D point vortex systems.
 Yet, for simplicity here I will only present a general discussion for $N$ body systems in 3D.
 The applications to the Coulomb systems will be published jointly with Carlo Lancellotti 
\cite{KieLan08}.

 Finally, to simplify notation I absorb normalizing factors such as \Ref{eq:ZUSTANDSSUMME} 
into the measures by writing $\underline\delta(\cdots)$ for normalized delta measures, and I
write the sets on which the $\underline\delta$'s are supported as subscripts.

\section{Folklore}

  Without attempting to give a verbatim quotation from any particular book, 
I believe the following statement (S) quite accurately captures the essence of similar 
statements in the pertinent literature:\hfill

\smallskip
\noindent 
(S) ``{\it An isolated classical $N$-body system with Hamiltonian $H$ and energy 
$E$ will spend most of its time in thermal equilibrium. 
 In statistical mechanics, this equilibrium state is given by Boltzmann's stationary
ergodic ensemble measure
\begin{equation}
\dd\mu =  \underline\delta_{\{H=E\}}(\dd^{3N}\!p\,\dd^{3N}\!q),
\label{eq:theERGODE}
\end{equation}
unless additional constants of the motion exist; in this case there is a 
stationary probability measure for each ergodic component ${\cal M}$, given by 
\begin{equation}
\dd\mu 
=  \underline\delta_{\cal M}(\dd^{3N}\!p\,\dd^{3N}\!q),
\label{eq:subERGODE}
\end{equation}
where ${\cal M} ={\cap_{k=0}^K \{I_k =C_k\}}$, 
the $I_k$ are the isolating integrals, with $I_0\equiv H$,
the $C_k$ their values, and $\dd^{3N}\!p\,\dd^{3N}\!q$ 
stands for Liouville measure on phase space.}''

  Since evaluating \Ref{eq:theERGODE} in one of the usual ways (that is approximately, 
in most cases) will yield very satisfactory agreement with the empirical data almost anytime 
we inquire into the thermal equilibrium properties of a many body system which can be 
described classically, and since the general empirical relations of thermal equilibrium
are recovered rigorously from \Ref{eq:theERGODE} in some 
appropriate thermodynamic limit $N\to\infty$ dictated by the Hamiltonian, most of us, 
presumably, do not take any offense at reading statements such as (S), although
we should.
  Indeed, in the line above \Ref{eq:theERGODE}, (S) makes the unfortunate 
suggestion that the {\it macroscopic thermal equilibrium state} of an {\it individual} 
physical system should be identified with a {\it stationary probability measure} over 
an {\it ensemble} of such systems, and this leads to some grossly incorrect conclusions
in all but the most favorable circumstances.

  What's more, (S) is conceptually absurd. 
  For instance, many invoke information theory to justify
probability measures such as \Ref{eq:theERGODE} and \Ref{eq:subERGODE}, 
becoming subjective, and then (S) violates what should be enshrined as one 
of the 10 commandments of physics:

\smallskip
\centerline{$\cal THOU\ SHALT\ NOT\ MIX\ ONTOLOGY\ WITH\ EPISTEMOLOGY!$\footnote{Read:
  Do not confuse what a system does with what you can know about it! $\phantom{phantom}$
  (The negation of this commandment is being worshiped in the orthodox quantum community,
  which seems exclusively concerned with observation and, most recently, information.
  But I don't want to take off on this tangent here.
  I have expressed my views in \cite{Kie08}.)}}

\smallskip

  But even if the ensemble probabilities are interpreted in an objective manner 
the identification of the equilibrium macrostate of any {\it individual} physical system 
with a stationary {\it ensemble} probability measure is conceptually absurd, although
in favorable cases  when the overwhelming majority of systems in the ensemble have 
essentially the same macrostate then the results computed with this macrostate
agree quantitatively with the results computed with the ensemble measure.
  Clearly, in such favorable cases some {\it law of large numbers} holds for the 
ensemble, i.e. if one formally let's $N\to\infty$ {\it the empirical statistics
of almost all members of the ensemble agrees with its ensemble mean}. 
  Boltzmann showed that such a law of large numbers holds for the perfect gas, assuming that
$H$ is the only conserved quantity and using the ensemble measure \Ref{eq:theERGODE}.
  This prototype result by Boltzmann may have suggested the kind of thinking expressed in (S);
all the same, (S) {\it is} absurd, and while masked by quantitative insignificance whenever a 
law of large numbers holds for the ensemble, the absurdity is there to stay and becomes 
manifest whenever the ensemble as a whole does not feature a law of large numbers --- 
which is the case, for instance, at phase transitions of first order.
  Such phase transitions are defined quite sharply in the formalism of thermodynamics 
but not for \Ref{eq:theERGODE} when $N<\infty$, yet the situation is quite easy to understand: 
if roughly fifty percent of the ensemble members have ``spin $+1$,'' the other half have
 ``spin $-1$'' (to use a simple representative example to illustrate the idea), 
the ensemble mean is ``spin zero;'' yet, each member of the ensemble has either spin 
$+1$ or spin $-1$, not spin $\approx 0$.
  
  Incidentally, Boltzmann was well aware of the fact that the notion of a phase transition 
and other thermodynamic notions as well are not unambiguously defined in statistical mechanics
for any $N<\infty$, and he also understood that all these thermodynamic notions can be 
sharply recovered from statistical mechanics in some limit $N\to\infty$.
  To rigorously implement this idea took a little longer, until Ruelle, Fisher, Lebowitz, 
Lieb, Thirring and others published their seminal papers on the subject in the 60s and 70s 
of the previous century, see \cite{RuelleBOOK, ThirringBOOK}.
  The insight that the thermodynamic notions are not exact for $N<\infty$ yet rigorously 
recovered when $N\to\infty$ has lead to the suggestion that {\it statements like (S) have 
to be understood as implicitly referring to the (or some) ``thermodynamic limit'' $N=\infty$,
with a system's macrostate being identified in general not with the limit of the measure 
\Ref{eq:theERGODE} or \Ref{eq:subERGODE} itself but with the extremal ergodic measures
into which the limiting ensemble measure decomposes\footnote{The set of probability 
  measures on some measure space is convex, and every point in a convex set can be 
  written as a convex linear combination of the extremal points of the set.}
 --- the so-called pure states.}
  This avoids the spin-up vs. spin-down problem of the above example (and all problems
with similar dichotomies): the extremal ergodic measures are precisely the spin$-1$ and 
spin$+1$ measures.
  The pure states of the $N=\infty$ ensemble define ergodic sub-ensembles; a law of 
large numbers holds for these sub-ensembles though not for the full ensemble.
  But has this now eliminated the conceptual problems?

  Although the invocation of the limit $N\to\infty$ combined with the decomposition into 
extremal measures avoids the problems with dichotomies, the proposal to 
{\it define what is meant} by a (macro)state of an individual $N$-body system 
(as distinct from computing it to a high degree of precision) in terms of the limit 
$N\to\infty$ was once commented on by Wigner thus:\footnote{As told to me by Shelly Goldstein.
   Wigner commented on the suggestion by a distinguished seminar speaker that 
   the  measurement-induced collapse of the $n$-body wave function of the measured system,
   postulated in orthodox quantum mechanics, would perhaps result rigorously from the
   unitary Schr\"odinger evolution of the combined, $n+N$-body system in the limit 
   $N\to\infty$, just as sharp phase transitions do. 
   Clearly, the gist of Wigner's remark is also valid in our context.}
 ``[Name], you are a great man, but you are are not infinite.''
Wigner's point, translated into our setting, was that even granted 
the extremal ergodic measures of the $N=\infty$ ensemble were legitimate 
macrostates {\it of the $N=\infty$ system} (they are, as we shall see, though there 
is still a subtlety to be addressed),  
that still doesn't free us from the obligation to {\it define} a macrostate 
of a finite-$N$ system {\it without invoking the limit $N\to\infty$}. 
  Of course, having found a sensible definition of macrostate for the $N=\infty$ system, 
a finite-$N$ definition should coincide with the $N=\infty$ definition in the limit. 
  I will give a legitimate definition below.

  I now address the second distinction of importance to which I have alerted you 
by highlighting in the paragraph after (S): the distinction between the {\it equilibrium}
of an individual system and the {\it stationarity} of an ensemble measure.
  It is very easy to be confused about this distinction, {\it even if} you are not confused 
about the distinction between macrostate of an individual system and the state (i.e. 
probability measure) of an ensemble of such systems as  discussed above.
  Part of the confusion is just a matter of semantics. 
  With hindsight, Gibbs' choice of terminology, to call ensembles defined by 
\Ref{eq:theERGODE} or \Ref{eq:subERGODE} {\it statistical equilibrium ensembles}
 \cite{Gibbs}, while sensible, was not very fortunate.
  Another semantic item which contributes to the confusion is the fact that 
{\it thermodynamic equilibrium} of  an 
individual $N$-body system has become synonymous with its {\it statistical equilibrium}, 
which terminology is based on the same notions which are behind Gibbs' terminology for
a stationary ensemble of systems, only now applied to an individual system viewed as
an {\it ensemble of particles}.
  The provocative title of my presentation is deliberately chosen to put you up
against all the confusion right at the start: I insist, 
the title makes sense --- but which sense? 
  To find out we have now to confront the {\it serious confusion}, 
which is not at all a matter of semantics but belongs in a category described 
beautifully by Franz Kafka, who said ``To correctly grasp an issue and 
to misunderstand the same [issue] do not exclude one another.''
(Translated from a quotation in \cite{RuelleBOOK}.)

  A stationary ensemble measure by its very definition does not depend on time, and
this of course remains so also when taking the limit $N\to\infty$.
  The limiting measure has a unique convex decomposition into extremal measures, and
since the limiting measure is time-independent, so is its extremal decomposition. 
  If these extremal states represent the $N\to\infty$ limits of the possible 
macrostates of individual finite-$N$ systems in the ensemble, as we said earlier they
would, only time-independent macrostates can~be~obtained.

  Here then is something paradoxical: all the measures \Ref{eq:subERGODE} are stationary,
so what I just said applies to them.
  In particular, it applies to the special ensemble measure of the type \Ref{eq:subERGODE} 
which is the next best known after \Ref{eq:theERGODE}, namely the one for 
${\cal M} = \{H =E\}\cap \{J = L\}$
where the Hamiltonian $H$ and the $z$-component (say) $J$ of the angular momentum are
conserved, having values $E$ and $L$, respectively.
  Now when $L\neq 0$ the system must rotate, in fact rigidly as Landau and Lifshitz  
will tell you.
  But we know that there are situations where the rotating system is not rotationally
invariant, and then the macrostate of this system will be time-dependent, of the kind
$\cos(\omega t - \varphi)$, with $\omega$ the constant angular frequency of the rigid
rotation.
  For example, think of the triaxial Jacobi ellipsoids which describe rigidly rotating 
constant density self-gravitating fluids \cite{Chandra}.
  How does this all fit together?

  According to the folklore the paradox is resolved as follows:

\noindent
(S$^\prime$)
``{\it Since a stationary ensemble only yields stationary macrostates (read: equilibrium 
states), and since the rigidly rotating configurations with non-vanishing angular momentum 
rotate with a constant angular frequency, one has to apply the methods of equilibrium 
statistical mechanics} in a co-rotating frame {\it where the macrostate of the individual 
system is now stationary.}''

\noindent
 With regard to an inertial frame such rotating macrostates are usually called 
{\it rotating equilibria}.
  This prescription of how to apply the equilibrium statistical mechanics formalism
leaves a strange taste in one's mouth, for it avoids the answer of the puzzle by 
circumventing it; yet the described procedure has worked fine since its inception, 
also for me.
  
  After relegating the above paradox to the back of my mind for many years, it
finally came full back into my conscience after Carlo Lancellotti and I had begun to work 
on the rigorous derivation of  what the Lynden-Bells \cite{LBLBa,LBLBb} had called a 
``perpetually pulsating equilibrium.'' 
  They had discovered a curious Hamiltonian $N$-body system in three dimensional space 
with an unusual additional conservation law, besides the conventional ones for energy,  
momentum, and angular momentum.
  They managed to show, by studying the virial theorem, that the mean-square radial 
distance from the center of mass undergoes a sinusoidal motion for all $N$. 
  They had also carried out some computer simulations with moderate $N$ and looked at
the empirical statistics of velocities in a ``co-pulsating frame,'' finding evidence 
for a Maxwellian --- hence their terminology. 
  But when Carlo and I set out to apply the conventional strategy advocated in the 
folklore (S$^\prime$), i.e. transforming the problem into a co-moving frame, I soon 
grew dissatisfied with this approach for conceptual reasons. 
  The dynamics of the co-moving frame, while still simple, is already 
quite non-trivial because now the angular frequency of the rigid rotation is itself 
time-dependent (think of the familiar sight of an ice skater performing a pirouette, 
with her arms periodically extending from and retracting to her body).
  If we permit terms like ``rotating equilibrium'' and ``pulsating equilibrium,'' 
we certainly have to also allow ``rotating-pulsating equilibrium,'' as this is
what the Lynden-Bells' ``equilibrium'' does in general. 
  But what if someone discovers an even more curious Hamiltonian $N$-body system,
with a ``rotating-pulsating-twisting-bending-shearing-jerking-and-what-have-you
equilibrium'' --- where does it stop?
  Are we not diluting the meaning of the word ``equilibrium'' until the construction
becomes totally absurd?

  Somehow the folklore was totally off target.

\section{Epiphany}

  The crucial insight is that something  everyone knows to be true anyhow for the 
finite-$N$ measures \Ref{eq:subERGODE} in regard to the actual finite-$N$ microstates
is true --- by inheritance --- also for the macrostates of the actual individual systems 
in the ensemble, and remains true in the limit $N\to\infty$. 

  In a nutshell, \Ref{eq:subERGODE} lives on the $N$-body phase space, which consists
of the {\it generic} phase points of $N$-body systems.
  Any point in $N$-body phase space defines a possible microstate of an $N$-body 
system.
  In addition, there is a Hamiltonian flow on this generic $N$-point phase space.
  Generic points do not move, they are independent variables.
  What evolves in time is the {\it actual} point in phase space which 
identifies {\it the microstate} of the actual $N$-body system. 
  Its motion follows a particular flow line (if the Hamiltonian is time-independent)
picked out by the initial actual phase point.
  So while \Ref{eq:subERGODE} is stationary (with respect to the flow on $N$-body phase space),
the ensemble of actual $N$-body systems it represents consists of systems whose actual 
phase points in general will not be stationary. 
  Everyone knows that, and no one has a problem with that, even though some people
may prefer to say the same thing differently.

  Now, any mapping of the $N$-body microstates into $N$-body macrostates also induces
a mapping of the flow on the space of generic microstates into a corresponding ``flow'' on 
the space of generic macrostates. 
  Let's call this new ``flow'' the macroflow.
  Also the measure \Ref{eq:subERGODE} will accordingly be mapped into a stationary measure on 
the space of generic macrostates. 
  Let's call that induced ensemble measure the macrostate measure.
  If in the limit $N\to\infty$ the macrostate measure is supported only on such generic 
macrostates which are stagnation points for the limiting macroflow, then the limiting 
macrostates of the actual systems will be stationary too, and so the true finite-$N$ 
macrostates will be approximately stationary, with small fluctuations.
  It is in the nature of macrostates that they don't care about most of what is going on
at the microscopic level, so macrostates can be relatively quiet even if the underlying
microstates evolve violently.
  But if that limiting measure is supported on generic macrostates at which the limiting
macroflow is not stagnant, then the limiting macrostates of the actual systems will be 
dynamical, traversing an orbit in the subset of generic macrostates which form the support 
of the limiting stationary macrostate measure. 
  By the same token as before, the true finite-$N$ macrostates will exhibit similar 
large scale macroscopic motions.

  It should be perfectly clear by now that there is no conflict between the ensembles being 
stationary, implying that their extremal decompositions in the limit $N\to\infty$ are 
stationary as well, and the undeniable fact that the macrostates of individual actual
systems making up the ensemble will in general exhibit macroscopic evolution in time. 

  In the remaining sections I will carefully explain how this general story pans out
for the ensembles \Ref{eq:subERGODE}.
  I hope that my explanations are entirely correct, but I cannot rule out that I 
am still confused about this or that --- think of Kafka!
  As appropriate for this conference, the general story is worked out for systems 
with long-range Galilei-invariant pair interaction which I will specify in more 
detail below; yet, the next section and part of section 5 are valid for Hamiltonians $H$ 
with more general Galilei invariant pair interactions.
  I will not invoke transformations into any co-moving non-inertial frames whatever; 
however, a Galilei boost into an inertial frame in which the center of mass is at rest 
is of a different category and will be performed (though the formalism does not even 
require that!).
  Of course, some limit $N\to\infty$ will be invoked, though not to define 
macrostates of the finite-$N$ systems but to calculate them approximately. 
  Macrostates will be defined for finite-$N$ systems without invoking that limit.
  In fact, our first duty is to define what we mean by {\it microscopic} 
and {\it macroscopic states}, 
or {\it micro-} and {\it macrostates} for short.
 There is also an intermediate level frequently referred to as {\it mesoscopic}.

\section{Micro-, meso-, and macroscopic}

  Taken literally the term ``...-scopic'' means something like ``point of view,''
a very subjective notion, and we sure don't want to confuse what systems do
with how we look at them!
 Fortunately there is a different way of putting it, namely as the
objective structures produced by the $N$-body dynamics at the ``finest scales,'' at
``intermediate scales,'' and at ``large scales.''
 There is still some ambiguity as to what exactly is ``fine,'' ``intermediate,'' and
``large,'' and this ambiguity will not go away!
 Yet at least the reference to subjective notions which seem implied by ``scopic'' is gone.

 However, concepts must not be too arbitrary or they lose their significance.
 Since by ``state'' of a system one means a complete specification of the dynamical variables
of the system, it must evolve (to good approximation) according to an autonomous 
dynamics {\it at the scale under consideration}; cf. \cite{ShellyJoel}.

\subsection{Microstates of individual systems}

 Although the term {\it microscopic scale} does contain the ambiguity aluded to above,
there fortunately is a clear cut choice for what the {\it microscopic state} of a
Hamiltonian $N$-body system should be. 
  Our Hamiltonian systems consist of $N\geq 2$ interacting nonrelativistic Newtonian 
point particles in Euclidean $\Eset^3$ with generic momenta $\pVi\in\Rset^3$ and positions 
$\qVi\in\Rset^3$ in an arbitrarily chosen inertial frame $\Rset^3$.
 Thus, for our Hamiltonian $N$-body systems then any point
$(\pV_1,...,\pV_N,\qV_1,...,\qV_N)=X\in\Rset^{6N}$
represents a possible state of the $N$-body system, or a {\it generic state}.
 If $\qVi(t)$ and $\pVi(t)$, $i=1,...,N$, are the $N$ particles' actual position and 
momentum vectors at time $t\in\Rset$ in an inertial frame, then 
$(\pV_1(t),...,\qV_N(t))=X_t$ is the {\it actual state} at time $t$.
 This is the most detailed characterization possible for the Hamiltonian $N$-body system,
and therefore this is its {\it microstate}.
 This is the conventional definition.
 It is natural and unambiguous. 

 The disadvantage of this conventional definition of the microstate is that it is not
well tailored to our needs. 
 So instead we will work with a completely equivalent characterization of the microstate
of a Hamiltonian $N$-body system, equivalent in the sense that there will be a one-to-one
mapping between both types of objects.
 	We identify each point $(\pV_1,...,\qV_N) =X$
with a singular \emph{empirical ``density''} (a measure) on $\Rset^6$ as follows:
\begin{equation}
\Delta_X(\pV,\qV)
 = 
\sum_{i=1}^N \delta(\pV-\pV_i)\delta(\qV-\qV_i).
\label{genMICstate}
\end{equation}
 Obviously the map is bijective, i.e. $X \leftrightarrow\Delta_X$.
 Henceforth we will refer to \Ref{genMICstate} as the generic microstate of an $N$-body
system. 
 Note that \Ref{genMICstate} means physically the number density of particles at $q\in\Rset^3$
with momentum $p\in\Rset^3$.
 Since by $X_t$ we denote the actual point in phase space $\Rset^{6N}$ at time $t$, 
then for us $\Delta_{X_t}(\pV,\qV)$ is the actual microstate of the $N$-body system.

 For the sake of completeness, we remark that this concept of microstate extends also 
to an $\infty$ many particles system.
 Intuitively: countably infinitely many points in $\Rset^6$ have Lebesgue measure 0. 
 Even though the empirical measure $\sum_{i=1}^\infty \delta(\pV-\pV_i)\delta(\qV-\qV_i)$ is 
not even locally finite if $\infty$ many points are densely distributed in some domain 
$\subset\Rset^6$, one can make sense out of it by looking at it as 
the collection of its finite $N$ truncations equipped with total variation ($TV$) topology. 
 In $TV$ topology then there will be no convergence to a continuum function when $N\to\infty$,
so even though the $\infty$ many points may be dense in some domain, the topology is fine
enough to distinguish the state from a continuum state.
 However, we will not need this construction.

\subsection{Mesostates of individual systems}

 We now face an irreducible ambiguity: what exactly constitutes a 
``mesoscopic scale''? 
  There is no such thing as ``{\it the} mesostate'' of an $N$-body system
per se.
 Yet in statistical physics one usually invokes this terminology in the context of 
kinetic theory, when a system is described in terms of some continuum formalism
in $(\pV,\qV)$ space $\Rset^6$; think of Boltzmann's transport equation for a dilute gas,
or Vlasov's equations. 
 Accordingly we could be tempted to call any such continuum density function $f(\pV,\qV;t)$
of kinetic theory an ``actual mesostate,'' and the time-independent generic functions
the ``generic mesostates.''
 This is not quite adequate, though.
 The reason is that a {\it continuum} theory is of course even more detailed than 
an $N$-body theory, since $f$ needs to be specified at uncountably many points. 
 This is hardly a description that considers only scales sufficiently bigger than 
the microscopic one of the previous subsection. 

 This inadequacy of the continuum description is ameliorated by restricting the (positive)
continuum density functions $f(\pV,\qV)$ in a suitable way. 
 For instance one could work with piecewise linear, Lipschitz continuous
functions with Lipschitz constant $<C$ for some $C$, and which linear pieces have 
their epicenters at a fixed ``hypercubic'' lattice.
 This would be quite adequate.

 To make our life easier we are a little more cavalier and allow functions 
$f(\pV,\qV)$ which are uniformly Lipschitz continuous and uniformly iterated 
Lipschitz continuous. 
 Basically this says that such functions cannot be too steep,  and cannot oscillate 
on too small scales; and since they are non-negative and need to integrate to a finite
number, they cannot develop singularities when $(\pV,\qV)$ go to infinity.
 We denote this class of functions $\ME$. 
 It is understood that some uniform bound on the Lipschitz and iterated Lipschitz 
constants is chosen; they clearly restrict the possibilities,
but how exactly is a matter of choice and good judgement and part of the irreducible
ambiguity of any notion such as mesoscopic, or for that matter also of macroscopic.

 Next we associate to each microstate $\Delta_X$ a mesostate $f_X$. 
 This map may be but need not be bijective.
 To simplify the discussion, we work with the normalized empirical
densities $\ul\Delta_X = N^{-1}\Delta_X$ and understand henceforth
that any $f_X$ is normalized such as to integrate to 1 as well. 
 We now pick a suitable Kantorovich-Rubinstein metric $\dKR$ (which metrices
weak convergence of normalized measures) and define $f_X$
as the minimizer of the KR distance between $\ul\Delta_X$ and $\ME$, i.e. 
\begin{equation}
\dKR\left(\ul\Delta_X
\,,\,
f_X \right)
=\min
\{\dKR\left(\ul\Delta_X
\,,\,
f \right)|f\in\ME\}
\end{equation}
if the minimum exists; if it doesn't, since the infimum of the set on the r.h.s. is certainly 
strictly positive, one can stipulate some convenient rule to choose an $f_X$ among
those $f$ which differ from $\ul\Delta_X$ in KR distance by not more than the infimum
plus  $\epsilon$.
 If that sounds somewhat ambiguous, that's because it is, but such ambiguities are
acceptable given the inherent ambiguity in the notion of a ``mesoscopic scale.''

\subsection{Macrostates of individual systems}

  Of course, the same kind of irreducible ambiguity that haunts 
the ``mesoscopic scale'' also raises the question of what exactly 
constitutes a ``macroscopic scale;'' and again there is no objective
answer to this. 
  In the statistical physics literature, the main distinction between the 
``macroscale'' and the ``mesoscale'' is that in the former one now does not 
resolve the variations in $\pV$ space. 
  More to the point, a {\it macrostate} is usually the small collection of
fluid dynamical functions on $\qV$ space $\Rset^3$, like mass density, velocity field, 
and kinetic energy density field or pressure field.
  Such functions can be obtained from the mesostates by taking the first few 
moments over $\pV$ space. 

\section{Finite $N$}
%
\subsection{Dynamics}

  The dynamics of each system is governed by a Hamiltonian of the type
\begin{equation}
H^{(N)}(\pV_1,\qV_1;\dots;\pV_N,\qV_N)
:=
 \sum_{1\leq i\leq N}
{\textstyle{\frac{1}{2 m_i}}}\abs{\pV_i}^2
+
\sum\sum_{\hskip-.7truecm 1\leq i < j\leq N} 
e_i e_j V(\abs{\qV_i-\qV_j}),
\label{eq:HAMgeneral}
\end{equation}
where the $m_i>0$ are the inertial masses and the $e_i\in \Rset$ ``generalized charges.'' 
 The pair interaction potential $V(r)$ will be of long range and locally integrable to
admit a mean-field limit when $N\to\infty$, but many facts do not depend on all the 
details of $V$. 
 In particular, the following familiar conservation laws hold for a larger class of $V$
than just those allowing mean-field limits.
 Since we need the conservation laws to set up our ensembles, we collect them here.
\begin{lemma}
	Assume the pair interaction potential
$V$ satisfies   $V\in \Csp^2(\Rset_+,\Rset)$, guaranteeing
local existence and uniqueness of the dynamics.
	Then by the invariance of $H^{(N)}$ under time and space translations, 
as well as space rotations,
the isolating integrals of motion are: the Hamiltonian $H^{(N)}$ given in 
\Ref{eq:HAMgeneral},
the total momentum phase space function
\begin{equation}
\pVN^{(N)} (\pV_1,\qV_1;\dots;\pV_N,\qV_N)
:=
 \sum_{1\leq i\leq N} \pV_i,
\label{eq:mom}
\end{equation}
and the total angular momentum phase space function
\begin{equation}
\jVN^{(N)}(\pV_1,\qV_1;\dots;\pV_N,\qV_N)
:=
\sum_{1\leq i\leq N} \qV_i\times \pV_i.
\label{eq:angmom}
\end{equation}
\end{lemma}
 Lemma 1 exhausts the isolating integrals of motion w.r.t. the 
center-of-mass frames
which are shared by all Hamiltonian systems characterized by 
\Ref{eq:HAMgeneral} under the stated hypotheses on $V$.
	Additional isolating integrals for all $N\geq 2$ may exist for 
special cases of $V$, and we shall encounter an example later on.

	Moreover, on account of $H^{(N)}$ changing only by an additive 
constant under a Galilei boost, what is sometimes called the system's 
centroid and given by
$\frac{1}{M}\sum_{1\leq i\leq N} (m_i\qVi(t) - t\pV_i(t))$, 
is conserved, too;
here $M= \sum_{1\leq i\leq N} m_i$ is the total mass of the system,
which is independent of the choice of inertial frame.
	However, the centroid is not the evaluation of a $t$-independent 
function on $(\pV_1,\dots,\qV_N)$ space and therefore not an isolating
integral. 
  Yet, since by the Galilean invariance of Newtonian point mechanics any 
particular one of our $N$-body systems can always be described w.r.t. any 
of its center-of-mass frames (any co-moving inertial frame for which the center 
of mass is the origin of that frame), without loss of generality it suffices to
study $N$-body systems under the {\it holonomic scleronomous constraints}
$\pVN^{(N)} (\pV_1,\dots,\qV_N) =\zeroV$ and
$\cVN^{(N)}(\pV_1,\qV_1;\dots;\pV_N,\qV_N)  =\zeroV$,
where
\begin{equation}
\cVN^{(N)}(\pV_1,\qV_1;\dots;\pV_N,\qV_N) 
: = 
\frac{1}{M} \sum_{1\leq i\leq N} m_i\qVi.
\label{eq:CENTERofMASS}
\end{equation}
is the {\it center-of-mass phase space function}.
 As far as the stationary ergodic subensembles go that we introduce below, these 
center-of-mass-frame constraints can be treated on the same footing as the 
the true isolating integrals.

 Hence, to unify the notation, we include both the true isolating integrals
and the scleronomic constraint functions in a list of functions $I_k$, $k=0,1,...$, 
simply called {\it invariants}, with $C_k$ denoting the values they take.
 In particular, we {\it define}:
$I_0:=H^{(N)}$ with $C_0=E$,
$(I_1,I_2,I_3):=\pVN^{(N)}$ with $(C_1,C_2,C_3):=\zeroV$, 
$(I_4,I_5,I_6):=\jVN^{(N)}$ with $(C_4,C_5,C_6)=\lVN$, 
and 
$(I_7,I_8,I_9):=\cVN^{(N)}$ with $(C_7,C_8,C_9)=\zeroV$,
where $E\in\Rset$ and $\lVN\in\Rset^3$ are time-independent parameters.

  Incidentally, the phase space functions \Ref{eq:HAMgeneral}, \Ref{eq:mom}, \Ref{eq:angmom},
\Ref{eq:CENTERofMASS} are not all in involution; the invariants
$H^{(N)}$, $\pVN^{(N)}$, $|\jVN^{(N)}|^2$, $\jVN^{(N)}\cdot \eV$ are,
where $\eV$ is any fixed unit vector in space.
  Moreover, in any of our center-of-mass frames $\cVN^{(N)}$ is in
involution with $H^{(N)}$ but not with $\pVN^{(N)}$ and $\jVN^{(N)}$.
  This is not a problem, for we will characterise the ergodic submanifolds conveniently 
in such a way that we do not need to find a set of invariants in involution; however, it is
implicitly understood in the following that any additional invariants
$I_k$ with $k=10,...$ will not be redundant.
\begin{lemma}
	Let $K+1(\geq 10)$ be the number of invariants
admitted by $H^{(N)}$, with $V$ as in Lemma 1. 

	a) Let $\Mset_K$ denote the manifold
\begin{equation}
\Mset_K \equiv \left\{(\pV_1,\dots,\qV_N)\in\Rset^{6N}| 
			I_k=C_k, k=0,...,K\right\}.
\end{equation}
	Then $\Mset_K$ is invariant under the flow generated by the 
Hamiltonian $H^{(N)}$. 

	b) Let $\mu^{(N)}$ denote the singular measure supported on $\Mset_K$
given by
\begin{equation}
\dd\mu^{(N)} =  \delta_{\cap_{k=0}^K \{I_k =C_k\}}(\dd^{3N}\!p\,\dd^{3N}\!q).
\label{eq:generalMC}
\end{equation}
	Then \Ref{eq:generalMC} is invariant under the adjoint flow 
associated with $H^{(N)}$, i.e. \Ref{eq:generalMC} 
is a stationary weak solution of Liouville's equation 
\begin{equation}
\pdt \mu + \{\mu, H^{(N)}\} =0.
\label{eq:LIOUVILLE}
\end{equation}
\end{lemma}

	So far, the conditions for Lemmas 1 and 2 also cover
Hamiltonian systems of type  \Ref{eq:HAMgeneral} with physically relevant
pair interactions, such as 
$V(x) = K /x$ with $e_i\in \Rset$ (Coulombian electricity)
and
$V(x) = - K /x$ with $e_i\in \Rset_+$ (Newtonian gravity),
as well as
 $V(x) = K (x^{-12} - x^{-6})$ with $e_i\in\Rset_+$
(Lennard--Jones molecular pair potential).
	Unfortunately, neither of these Hamiltonian systems is self-confining,
and the jury is still out on the question of the generic global existence of 
the dynamics in the gravitational and electrical $N$-body problems.
	We now impose stronger conditions on $V$ which guarantee
global existence of the dynamics and self-confinement of the system
of particles.

\begin{proposition}
	Let $K+1(\geq 10)$ be the number of invariants
admitted by $H^{(N)}$, with $V$ satisfying the hypotheses in Lemma 1. 

	a) If in fact $V\in \Csp^2(\Rset_+,\Rset_+)$ with 
$-e_ie_jV^\prime(r)\stackrel{r\downarrow 0}{\longrightarrow} \infty$ 
for all $i,j$, or $V\in \Csp^2(\ol{\Rset_+},\Rset_+)$ with 
$V^\prime(r)\stackrel{r\downarrow 0}{\longrightarrow} 0$,
and in either case $\limsup_{r\to\infty}|V^\prime(r)|/r\leq C$,
then the dynamics exists globally in time.

	b) If in addition to the hypotheses in a)
we have $\lim_{r\to\infty}e_ie_jV(r) =+\infty$
for all pairs $i,j$, then \Ref{eq:generalMC} is a 
finite measure.
\end{proposition}

	Henceforth we assume that $V$ satisfies the hypotheses of
Proposition 1b).
	Note that when $N>2$ this implies that either $e_i>0 \ \forall i$
or $e_i<0 \ \forall i$.
      
\subsection{Statistics}

  In the section ``Folklore'' I spoke several times of ``a law of large numbers''
as $N\to\infty$. 
  This is related to, but not to be confused with, a familiar law of large numbers which
holds (under mild conditions) for ensembles of i.i.d. finite-$N$(!) systems.

	Denoting normalized \Ref{eq:generalMC} by $\ul\mu^{(N)}$, i.e. 
$\int\dd\ul\mu^{(N)} = 1$, this can be interpreted 
as the single-system a-priori probability measure for a Newtonian 
$N$-body system in phase space $\Rset^{6N}$ to be at $X$ if all that is given about the
system is its Hamiltonian $H^{(N)}$ with energy $E$, its total angular momentum $\lVN$, 
its center of mass $=\zeroV$, its total momentum $=\zeroV$, and the values 
$C_k$ for whatever other invariants there are; 
with $H^{(N)}$ also the number $N$ of particles and total mass $M$ are given.
	Thus, if $\mathscr{B}$ denotes the Borel sets of $\Rset^{6N}$, then
 $(\Rset^{6N},\mathscr{B},\ul\mu^{(N)})$ is our single-system probability space.
	The conventional \emph{microstate} of a single system is now a \emph{random vector}
$\cX = (\cP_1,\dots,\cP_N,\cQ_1,\dots,\cQ_N)\in \Rset^{3N}\times\Rset^{3N}$, 
the specification of which amounts to giving exactly the positions and 
momenta of each particle in the $N$-body system. 
	If $B\subset\Rset^{6N}$ is a Borel set, then the probability
of $\cX$ being in $B$ is 
\begin{equation}
\Prob\left(\cX\in B \right) = \ul\mu^{(N)}(B).
\end{equation}
	Clearly, $\Prob\left(\cX\in B \right) = 0$ unless 
$B\cap \Mset_K \neq \emptyset$.

	The Boltzmann ergodic subensemble associated with 
\Ref{eq:generalMC} is an infinite family of i.i.d. random vectors $\cX^{(j)}$,
each with a-priori distribution \Ref{eq:generalMC}, and each one of which 
represents a Newtonian $N$-body system which is governed by the same 
Hamiltonian $H^{(N)}$ and has the same energy $E$, center of mass $=\zeroV$,
total momentum $=\zeroV$, total angular momentum $\lVN$, and same 
values $C_k$ for whatever other invariants there are; 
of course, each system also has the same number $N$ of particles and total
mass $M$.
	Any ordered set of $\ell$ copies of $\cX$ is a  random vector
$(\cX^{(1)},...,\cX^{(\ell)})\in \Rset^{\ell 6N}$ with joint distribution 
$\ul\mu^{(N)}{}^{\times \ell}$.
	For this family of i.i.d. systems the conventional
weak law of large numbers (WLLN) holds. 

	We are going to state the WLLN in the form adapted to 
our discussion of the concept of a system's ``microstate'' as $\Delta_X(\pV,\qV)$.
	Accordingly, $\cX$ is identified with a 
singular-empirical-density-valued random variable $\Delta_{\cX}(\pV,\qV)$,
and the measure \Ref{eq:generalMC} is identified with a
single-system a-priori probability measure $\tilde\mu^{(N)}$ on the convex set 
of all  singular densities (i.e. measures) on $\Rset^6$.
	Clearly, $\tilde\mu^{(N)}$ is supported only on the set of all singular
densities of the type $\Delta_X(\pV,\qV)$.
	Boltzmann's ergodic ensemble is therefore identical to an infinite 
family of i.i.d. random densities $\Delta_{\cX}(\pV,\qV)$, each with a-priori
distribution $\tilde\mu^{(N)}$.
	By convexity, the \emph{empirical mean} of a size-$\ell$ sample of 
i.i.d. empirical densities, 
\begin{equation}
\ol{\Delta_X}^{(\ell)}(\pV,\qV) 
= \frac{1}{\ell}\sum_{j=1}^\ell \Delta_{X^{(j)}}(\pV,\qV)
\end{equation}
is again a density of the same total measure $=N$.
	For the normalized measure $\ul\mu^{(N)}$ defined by \Ref{eq:generalMC}, denote by
$\ul\mu_{1,i}^{(N)}(\dd^3p\dd^3q)$ its first marginal measure for the $i$-th 
particle variables $(\pV_i,\qV_i)$, evaluated at $(\pV,\qV)$.
	Recall (see, e.g. \cite{DurrettBOOK, DudleyBOOK}) that a family of 
probability measures $\nu_n$ on $\Rset^d$ is said to converge 
weakly to $\nu$ if $\int g(x)\dd\nu_n\to\int g(x)\dd\nu$ for 
all bounded continuous functions $g\in \Csp^0_b(\Rset^d)$. 
	Recall furthermore (e.g. \cite{DudleyBOOK}) that weak convergence
is metrized by some Kantorovich-Rubinstein metric $\dKR$. 
	We are now ready to state the  weak law of large numbers.

\begin{theorem} 
\label{thm:WLLNa}
	For Boltzmann's ergodic ensemble of i.i.d. random vectors $\cX$,
each with a-priori distribution \Ref{eq:generalMC}, we have the weak
law of large numbers,
\begin{equation}
\lim_{\ell\to\infty}
\Prob\left(\dKR\left(\ol{\Delta_{\cX}}^{(\ell)}(\pV,\qV)\dd^3p\dd^3q 
\,,\,
\textstyle{\sum_{i=1}^N} \ul\mu_{1,i}^{(N)}(\dd^3p\dd^3q)\right)
> \eps\right) =0.
\end{equation}
\end{theorem}

	We follow the probabilists custom and rephrase the WLLN shorter thus:
\begin{equation}
\ol{\Delta_{\cX}}^{(\ell)}(\pV,\qV)\dd^3p\dd^3q 
\stackrel{\ell\to\infty}{\longrightarrow }
\sum_{i=1}^N \ul\mu_{1,i}(\dd^3p\dd^3q) 
\qquad\mathrm{in} \quad \mathrm{probability}.
\end{equation}

	So far all the particles in an individual $N$-body
system may have different masses and charges.\footnote{Physical examples of
	large $N$-body systems with a distribution of particle masses are
	stellar clusters.}
	Many systems of interest in nature do consist of only a handful of species, 
and the  prototype systems to study are one-specie systems with
$m_i=m>0$ and $e_i=e\in\Rset$ for all $i$ in \Ref{eq:HAMgeneral}, so the Hamiltonian
becomes
\begin{equation} 
H^{(N)} (\pV_1,\dots,\qV_N)
=
 \sum_{1\leq i\leq N}
{\textstyle{\frac{1}{2m}}}\abs{\pV_i}^2
+
\sum\sum_{\hskip-.7truecm 1\leq i < j\leq N} e^2V(\abs{\qV_i-\qV_j}).
\label{eq:HAM}
\end{equation}	
         For such single specie systems the WLLN simplifies to the following.
\begin{theorem}
\label{thm:WLLNb}
	For the Hamiltonian \Ref{eq:HAM}, all the 
first marginal measures are identical due to the permutation 
invariance of $H^{(N)}$ which is inherited by the product over all the
isolating invariants in \Ref{eq:generalMC}.
	Thus,
$\ul\mu_{1,i}^{(N)}(\dd^3p\dd^3q) = \ul\mu_{1}^{(N)}(\dd^3p\dd^3q)$ for all $i=1,...,N$.
	Accordingly, as $\ell\to\infty$, 
\begin{equation}
\ol{\Delta_{\cX}}^{(\ell)}(\pV,\qV)\dd^3p\dd^3q \longrightarrow 
N\ul\mu_{1}^{(N)}(\dd^3p\dd^3q)
\label{eq:WLLN}
\end{equation}
in probability.
\end{theorem}
 Note that no limit $N\to\infty$ is involved at this level.

\section{The limit $N\to\infty$}

  {From} now on our Hamiltonian will be \Ref{eq:HAM}.
  In addition, we are interested now only in those Hamiltonians \Ref{eq:HAM} which 
permit a mean-field continuum limit $N\to\infty$.
	Therefore, in addition to the hypotheses of Proposition 1b, 
from now on we assume also that $V$ satisfies 
$V\circ\abs{\,.\,} \in \Lsp^{1}_{\mathrm{loc}}(\Rset^3)$, and we also ask that
$\lim_{r\to\infty}V(r)/r=\infty$.

\subsection{Dynamics of microstates}

	Having identified points $X$ on $\Mset_K\subset\Rset^{6N}$ with 
empirical densities $\Delta_{X}(\pV,\qV)$, we note that
the Hamiltonian flow on $\Mset_K$ which transports any point
$X_0$ on $\Mset_K$ at time $0$ to a corresponding point $X_t$ on $\Mset_K$ 
at time $t$  also defines the dynamics of the empirical densities
$\Delta_{X}(\pV,\qV)$; namely, the initial empirical density
$\Delta_{X_0}(\pV,\qV)$ evolves into $\Delta_{X_t}(\pV,\qV)$ at time $t$.
	Under suitable assumptions the dynamics of the normalized empirical
densities $\ul\Delta_{X_t}(\pV,\qV)$ converges with $N\to\infty$ to a limiting 
dynamics of normalized continuum densities $f(\pV,\qV;t)$.

	In the following, we attach a superscript $^{(N)}$ to $\Mset_K$
and $X\in \Mset_K$ to indicate that $\Mset_K=\Mset_K^{(N)}$
and $X=X^{(N)}$ change with $N$.
	We also write $H^{(N)}(X^{(N)})$ for $H^{(N)}(\pV_1,...,\qV_N)$,
and similarly for the invariants $\pVN^{(N)}$, $\cVN^{(N)}$, and $\jVN^{(N)}$.

	We are interested in a continuum limit in the sense that the
particle positions $\qV_i$ (we should write $\cQ_i$) converge in distribution 
as $N\to\infty$, i.e. (at time $t=0$) we want 
$\int_{\Rset^3}\ul\Delta_{X^{(N)}_0}(\pV,\,.\,)\dd^3p
\stackrel{N\to\infty}{\longrightarrow}
\rho(\,.\,;0)\in (\Lsp^1_{+,1}\cap C^0_b)(\Rset^3)$ weakly
(metriced by some Kantorovich-Rubinstein metric $\dKR$).
	We are only interested in those $\rho$ for which
$\int\!\!\int V(\abs{\qV-\qV'})\rho(\qV;t)\rho(\qV';t)\dd^3q\dd^3q'<\infty$,
where $\int\!\!\int$ is taken over ${\Rset^3\times\Rset^3}$.
	We also want the kinetic and potential energies to scale comparably 
with $N$ for large $N$. 
	Since only $N$ summands contribute to the kinetic energy 
while the potential energy $\propto N^2$ (for $N\gg 1$) when
$\int_{\Rset^3}\ul\Delta_{X^{(N)}}(\pV,\qV)\dd^3p \to \rho(\qV)$ in 
the stipulated sense (suppressing $t$ temporarily), to achieve equal
$N$-scaling of kinetic and potential energies the rescaled particle 
momentum vectors $N^{-1/2}\pV_i$ must converge in 
distribution, so that 
$N^{-2} \sum_{1\leq i\leq N} \abs{\pV_i}^2 
\stackrel{N\to\infty}{\longrightarrow} 2 \vepskin$.
	Thus we ask that
$N^{3/2}\ul\Delta_{X^{(N)}}(N^{1/2}\pV,\qV)
	\stackrel{N\to\infty}{\longrightarrow}  
		f(\pV,\qV)\in (\Lsp^1_{+,1}\cap C^0_b)(\Rset^6)$
weakly, such that 
$N^{-2} H^{(N)}(X^{(N)})\stackrel{N\to\infty}{\longrightarrow} \pzcE(f)
= \veps <\infty$, 
where
\begin{eqnarray}
\pzcE(f)\!\! &=&\!\!  \frac{1}{2m}\iint |\pV|^2 f(\pV,\qV)\dd^3p\dd^3q 
\nonumber\\
&&\!\! + \frac{e^2}{2}\iint\!\!\!\iint
 V(\abs{\qV-\qV'})f(\pV,\qV)f(\pV',\qV')\dd^3p\dd^3q \dd^3p'\dd^3q'\,,
\label{eq:eOFf}
\end{eqnarray}
is the ``energy of $f$.''
	It then follows also that\footnote{The vanishing of 
		$\pzcQV(f)$ and $\pzcPV(f)$ follows of course from
		the fact that each $X^{(N)}$ is picked from a $\Mset_K^{(N)}$, 
		so that necessarily $\cVN^{(N)}(X^{(N)}) = \zeroV$ and
		$\pVN^{(N)}(X^{(N)}) = \zeroV$ for all $N$.}
$\cVN^{(N)}(X^{(N)})\stackrel{N\to\infty}{\longrightarrow}\pzcQV(f) = \zeroV$, 
$N^{-3/2}\pVN^{(N)}(X^{(N)})\stackrel{N\to\infty}{\longrightarrow}\pzcPV(f)=
\zeroV$,
and
$N^{-3/2}\jVN^{(N)}(X^{(N)})\stackrel{N\to\infty}{\longrightarrow}\pzcJV(f)=
\lV$, 
where 
\begin{equation}
\pzcQV(f) = \iint \qV f(\pV,\qV)\dd^3p\dd^3q
\label{eq:qOFf}
\end{equation}
\begin{equation}
\pzcPV(f) = \iint \pV f(\pV,\qV)\dd^3p\dd^3q
\label{eq:pOFf}
\end{equation}
\begin{equation}
\pzcJV(f) = \iint \qV\times \pV f(\pV,\qV)\dd^3p\dd^3q
\label{eq:jOFf}
\end{equation}
are the ``center of mass ---,'' ``momentum ---,'' and 
``angular momentum of $f$,'' respectively.
	The functionals $\pzcJV(f)$, $\pzcPV(f)$, $\pzcQV(f)$ are
well-defined for $f\in \Lsp^1_{+,1}$ whenever $\pzcE(f)<\infty$
thanks to our hypotheses on $V$.

	We are now ready to state our first dynamical mean-field limit theorem.
	It is understood throughout that $V$ in  \Ref{eq:HAM} 
satisfies the stipulated hypotheses.
	Recall that the list of invariants $\{I_0,...,I_9\}$ is common
to all Hamiltonian systems with a Hamiltonian of type \Ref{eq:HAM}.
\begin{theorem}
\label{thm:Vlasov}
	Let $H^{(N)}_9$ denote any Hamiltonian of type \Ref{eq:HAM} 
for which the list of invariants $\{I_0,...,I_9\}$ exhausts the 
invariants.
	Let
\begin{equation}
\Asp_9 = \{f \in \Lsp^1_{+,1}(\Rset^6): \pzcE(f) <\infty,
	|\pzcPV(f)|<\infty, |\pzcQV(f)|<\infty, |\pzcJV(f)|<\infty\}
\label{eq:Anine}
\end{equation}
denote the set of admissible densities on $\Rset^6$.
    Let $\{t\mapsto X^{(N)}_t\}_{N=2}^\infty$ denote a sequence of solutions 
of the Hamiltonian dynamics on $\Mset_9^{(N)}$, $N=2,3,...$, 
generated by $H^{(N)}_9$.
	For the initial data, assume that 
$N^{3/2}\ul\Delta_{X^{(N)}_0}(N^{1/2}\pV,\qV)
	\stackrel{N\to\infty}{\longrightarrow} f(\pV,\qV;0)
\in \Asp_9$ weakly.
	Then 
$N^{3/2}\ul\Delta_{X^{(N)}_{N^{-1/2}t}}(N^{1/2}\pV,\qV)
	\stackrel{N\to\infty}{\longrightarrow}f(\pV,\qV;t)
\in \Asp_9$ weakly, where $f(\pV,\qV;t)$ is a weak solution to the Vlasov 
equation
\begin{equation}
\pdt f + m^{-1}\pV\cdot\nabla f 
- \nabla \left(e^2V* {\textstyle\int} f\dd^3p\right)\cdot\pdp f =0
\label{eq:Vlasov}
\end{equation}
with initial data $f(\pV,\qV;0)$.
	Moreover,
$N^{-2} H^{(N)}(X^{(N)}_{N^{-1/2}t})\to \pzcE(f(\,.\,;t)) = \veps$, 
$\cVN^{(N)}(X^{(N)}_{N^{-1/2}t})\to \pzcQV(f(\,.\,;t)) = \zeroV$, 
and
$N^{-3/2}\pVN^{(N)}(X^{(N)}_{N^{-1/2}t})\to \pzcPV(f(\,.\,;t)) = \zeroV$, 
and 
$N^{-3/2}\jVN^{(N)}(X^{(N)}_{N^{-1/2}t})\to \pzcJV(f(\,.\,;t)) = \lV$, 
with $\veps$ and $\lV$ independent of time.
\end{theorem}

	Beside the invariants $\pzcI_k$ inherited from the
isolating invariants $I_k$, the Vlasov dynamics conserves 
an infinitude of so-called Casimir functionals of $f$.

\begin{theorem}
\label{thm:Casimir}
	Let $C:\ol{\Rset_+}\to\Rset$ be a continuous function for which
$\pzcC[f(\,.\,;0)]
 = \int\!\!\int C(f(\,.\,;0)) \dd^3p\dd^3q$ exists for $f\in\Asp_9$.
	Let $t\mapsto f(\,.\,;t)$ solve \Ref{eq:Vlasov} with
initial data $f(\,.\,;0)$.
	Then $\pzcC[f(\,.\,;0)] = \pzcC[f(\,.\,;t)]$ for all $t$.

In particular, among the Casimir functionals of \Ref{eq:Vlasov} are:

\noindent
the ``normalization of $f$,''\footnote{Of course, 
		$\pzcN(f)=1$ for $f\in \Asp_9$.}
\begin{equation}
\pzcN(f) = \iint f(\pV,\qV)\dd^3p\dd^3q\,,
\label{eq:mOFf}
\end{equation}
the ``entropy of $f$,''
\begin{equation}
\pzcS(f)
 = -\iint f(\pV,\qV) \ln f(\pV,\qV)\dd^3p\dd^3q\,.
\label{eq:sOFf}
\end{equation}
\end{theorem}

	If $H^{(N)}$ given in \Ref{eq:HAM} is not of type $H^{(N)}_9$, then
additional invariants $I_k$ ($k=10,...,K >9$) exist (because the list
$(I_1,...,I_9)$ is shared by all $H^{(N)}$ given by  \Ref{eq:HAM}.
	The Lynden-Bells' Hamiltonian is of type $H^{(N)}_{10}$ for which 
the obvious generalization of Theorem \ref{thm:Vlasov} holds, while Theorem
\ref{thm:Casimir} continues to hold unchanged. 

\subsection{The support of the stationary ensemble measures}

	We begin by stating our mean-field limit theorem for $\ul\mu^{(N)}$; 
or rather: $\tilde\mu^{(N)}$.

\begin{theorem}
\label{thm:maxENTROPYvp}
	Let $H^{(N)}=H^{(N)}_9$ as in Theorem \ref{thm:Vlasov}, set
$E=N^2\veps$ and $\lVN =N^{3/2}\lV$, with $\veps >0$ and $\lV$ fixed,
viable data.
	Let $\tilde\mu^{(N)}$ be given in \Ref{eq:generalMC} with $K=9$.
	Then 
\begin{equation}
\lim_{N\to\infty} \tilde\mu^{(N)} =\tilde\mu
\end{equation}
exists, and $\tilde\mu$ is a stationary measure for the adjoint dynamics
inherited from $H^{(N)}$ in the limit $N\to\infty$.
	Furthermore, $\tilde\mu$ is supported on the set of maximizers of the 
entropy functional $\pzcS(f)$, given in \Ref{eq:sOFf}, taken from the 
set of trial densities
\begin{equation}
\Tsp_9 =
 \{f \in \Lsp^1_{+,1}(\Rset^6): \pzcE(f) = \veps,
	\pzcPV(f)=\zeroV, \pzcQV(f)=\zeroV, \pzcJV(f)=\lV\}.
\label{eq:TRIALnine}
\end{equation}
	If ${f_\bullet}(\pV,\qV)$ is a maximizer of $\pzcS(f)$ over the set 
$\Tsp_9$, then ${f_\bullet}(\pV,\qV)$ satisfies the fixed point 
(Euler--Lagrange) equation
\begin{equation}
{f_\bullet}(\pV,\qV) = e^{-1+
	\lambda_\pzcN +
	     \lambda_\pzcE \left(
\frac{1}{2m}{|\pV|^2}+ e^2(V*\rho_\bullet)(\qV)
		  \right) +
	\lambdaV_\pzcJ \cdot \qV\times\pV +
	\lambdaV_\pzcP \cdot \pV +
	\lambdaV_\pzcQ \cdot \qV}
\label{eq:fixPOINT}
\end{equation}
where $\rho_\bullet=\int\!{f_\bullet}\:\!\dd^3p$ and where
$\lambda_\pzcN$, $\lambda_\pzcE$, $\lambdaV_\pzcJ$, $\lambdaV_\pzcP$,
$\lambdaV_\pzcQ$ are the Lagrange multipliers for the constraints
$\pzcN(f)=1$, $\pzcE(f)=\veps$, $\vec\pzcJ(f)=\lV$, $\vec\pzcP(f)=\zeroV$,
and $\vec\pzcQ(f)=\zeroV$.
\end{theorem}

	We are primarily concerned with the support of $\tilde\mu$.
	The interesting question of how $\tilde\mu$ varies over its support 
will be touched upon but briefly.

	In addition to the maximizers of  $\pzcS(f)$ there
may be other solutions of \Ref{eq:fixPOINT}, yet only those solutions 
${f_\bullet}$ for which $\pzcS({f_\bullet})=\max$
over $\Tsp_9$ are in the support of $\tilde\mu$.
	All solutions are critical points for $\pzcS(f)$.
	Independently of whether a solution of \Ref{eq:fixPOINT} 
maximizes $\pzcS({f_\bullet})=\max$ over $\Tsp_9$ or not the following holds. 

\begin{theorem}
\label{thm:STANDARDfixPOINT}
	Every solution of the fixed point equation \Ref{eq:fixPOINT}
satisfying the constraints $\pzcN(f_\bullet)=1$, 
$\vec\pzcP(f_\bullet)=\vect{0}$,
$\vec\pzcQ(f_\bullet)=\vect{0}$, and
$\vec\pzcJ(f_\bullet)=\lV$ 
factors into a product of a 
\emph{locally (at $\qV$) shifted Maxwellian} on $\pV$ space and a
purely space-dependent \emph{Boltzmann factor} with 
$\lambdaV_\pzcP=\vect{0}$, 
$\lambda_\pzcE = -1/\kB T<0$,
and $\lambdaV_\pzcJ = \omegaV/\kB T$, 
where $T$ is the temperature and $\omegaV$ the 
angular frequency vector of rotation of the dynamical system; thus
\begin{equation}
{f_\bullet}(\pV,\qV) 
= 
\left(\frac{1}{2\pi m\kB T}\right)^{\frac{3}{2}}
\exp\left(-\frac{1}{2m\kB T} \left|\pV-m\omegaV\times\qV\right|^2\right)
\rho_\bullet(\qV)
\label{eq:fixPOINTfactoredSTANDARD}
\end{equation}
with $\rho_\bullet$ satisfying the fixed point equation on $\qV$ space
\begin{equation}
\rho_\bullet(\qV)
=
\frac{\exp\left(-\frac{1}{\kB T}\left[{e^2}(V\!*\!\rho_\bullet)(\qV)
- \frac{m}{2}\left|\omegaV\times\qV \right|^2\right]
 + \lambdaV_\pzcQ \cdot \qV\right)}
{\int_{\Rset^3}\exp\left(-\frac{1}{\kB T}\left[
{e^2}(V\!*\!\rho_\bullet)(\qV')
- \frac{m}{2}\left|\omegaV\times\qV' \right|^2\right]
 + \lambdaV_\pzcQ \cdot \qV'\right)\dd{q'}}
\label{eq:fixPOINTrhoSTANDARD}
\end{equation}
	Here we also eliminated the Lagrange multiplier $\lambda_\pzcN$ 
using $\pzcN(f_\bullet)=1$.
\end{theorem}

	Theorems \ref{thm:maxENTROPYvp} and \ref{thm:STANDARDfixPOINT}
hold for all Hamiltonian systems with Hamiltonian of type $H^{(N)}_9$, which
covers -- presumably -- the majority of models with Hamiltonian given by 
\Ref{eq:HAM} satisfying the hypotheses of Proposition 1b together with those
stipulated at the beginning of the section on $N\to\infty$.
 They are proven similarly as in \cite{KieLebLMP}.
	It is straightforward to generalize 
Theorems \ref{thm:maxENTROPYvp} and \ref{thm:STANDARDfixPOINT} to the
Hamiltonians given by \Ref{eq:HAM} which are of type $H^{(N)}_{10}$,
which includes the Lynden-Bells' Hamiltonian.
 In that case the entropy maximizer factors into a product of a 
\emph{locally (at $\qV$) shifted Maxwellian} on $\pV$ space
and a purely space-dependent \emph{Boltzmann factor}.
  The shifted Maxwellian in \Ref{eq:fixPOINTfactoredSTANDARD} is known as
a ``rotating Maxwellian;'' in the case of the Lynden-Bells' Hamiltonian we find a
``rotating-dilating Maxwellian.''

\section{Thermodynamics [sic!]}
	
	By Theorem \ref{thm:STANDARDfixPOINT}, w.r.t. Kantorovich-Rubinstein topology
(metrizing weak convergence of measures) every solution of the fixed point equation 
\Ref{eq:fixPOINT} represents the microscopic density function on single particle 
phase space of an individual $N=\infty$ many particles system which 
\emph{locally} at each $\qV$ is in thermal equilibrium w.r.t. an 
instantaneously co-rotating frame at $\qV$.
	A forteriori this is true also for the maximizers of $\pzcS(f)$ 
over $\Tsp_9$.
	However, globally a solution to \Ref{eq:fixPOINT} --- and therefore
in particular any maximizer of $\pzcS(f)$ over $\Tsp_9$ ---
may be a stationary solution of \Ref{eq:Vlasov} but need not be.
	If a solution to \Ref{eq:fixPOINT} is a stationary
solution of \Ref{eq:Vlasov}, then it describes a system which as a whole
is either in a non-static stationary state or in a static state, depending 
on whether a non-vanishing velocity field 
$\ol\uV(\qV) =m^{-1}\int\!\pV{f_\bullet}(\pV,\qV)\:\!\dd^3p$
exists or not, respectively. 
	If a solution to \Ref{eq:fixPOINT} is not a stationary
solution of \Ref{eq:Vlasov}, then it is necessarily a snapshot of 
a whole connected orbit of densities which solve the genuinely time-dependent
Vlasov equation \Ref{eq:Vlasov}.
	Note though that all snapshot densities in such an orbit are necessarily 
maximizers of $\pzcS(f)$ over $\Tsp_9$, for the Vlasov dynamics preserves 
the isolating invariants as well as normalization and entropy.
  All this will become more dramatically so for the rotating-dilating
Maxwellians obtained for the Lynden-Bells' Hamiltonian of type  $H^{(N)}_{10}$.

  Incidentally, whenever these {\it microstates} of the system in the limit $N=\infty$
vary sufficiently mildly, then they coincide with the {\it mesostates} and 
map bijectively into the {\it macrostates}. 
 The upshot is: such solutions describe systems which in general are truly
thermo\emph{dynamical}!
	This leads us to tentatively propose the following terminology.

	The static maximizers of entropy  $\pzcS(f)$ over $\Tsp_9$ 
will be called \emph{thermostatic states}.
	The non-static stationary entropy maximizers
will be sanctioned \emph{thermostationary states}; these include some
(but not all) of those states which are frequently called 
``rotating equilibrium states.''
	The non-stationary  entropy maximizers
are genuinely \emph{thermodynamical states}.
        All this applies to the $N=\infty$ states. 

        The finite-$N$ meso- and macrostates for large enough $N$ are then
well-approximated by these $N=\infty$ states and can be categorized in these
terms as well. 
       This is the content of a law of large numbers as $N\to\infty$ over the
empirical statistics over the particles in an individual system. 
       Note that when there is a unique maximizer of entropy, then this law
of large numbers and the one stated earlier for finite $N$ will give essentially
the same results when $N$ is large enough, but only then.

\section{Closing remarks}

  In this presentation I have tried to explain what equilibrium statistical mechanics
is all about. 
  My title ``Statistical equilibrium dynamics'' summarizes this as concisely as I
could, in the following sense: Gibbs referred to Boltzmann's  stationary ergodic 
ensembles as being in ``statistical equilibrium'' (so these two words are bound 
also in my title), but the ``dynamics'' in my title refers to the meso- or macrostate 
of the individual members of such an ensemble, which as we saw, may very well 
carry out a nontrivial large scale dynamics even in the limit $N\to\infty$.
  Yet, at any instant of time the individual macrostate of the $N=\infty$ system
is a maximizer of the Boltzmann entropy function ($-H$ function) subject to the
the conservation laws and other scleronomic constraints.

 In this presentation I have restricted myself to the general picture, though I 
mentioned specific systems without descending into the details. 
 Those details will be given in a joint publication with Carlo Lancellotti \cite{KieLan08}.

{\textbf{Acknowledgment}} This work was supported in parts by the NSF under grants
DMS-0103808 and DMS-0406951. Any opinions expressed in this paper are entirely those
of the author and not those of the NSF. I thank Shelly Goldstein and Carlo Lancellotti 
for many valuable discussions. 
Finally, I thank the organizers, Alessandro Campa, Andrea Giansanti, Giovanna Morigi, 
and Francesco Sylos Labini, for the kind invitation to and the fantastic organization
of this workshop, and for their generous help and support. 



\end{document}